%% file: isit-v2.tex
\documentclass[10pt, twocolumn]{IEEEtran}

\usepackage{floatflt}

\usepackage{cite}

\usepackage{graphicx}

\usepackage{psfrag}

\usepackage{subfigure}

\usepackage{url}

\usepackage{stfloats}

\usepackage{amsmath}

\input{macros}

\newtheorem{thm}{Theorem}

\def\BibTeX{{\rm B\kern-.05em{\sc i\kern-.025em b}\kern-.08em
    T\kern-.1667em\lower.7ex\hbox{E}\kern-.125emX}}

\begin{document}

\title{Linear Finite-Field Deterministic Networks With Many Sources and One Destination}

\author{\authorblockN{M. Majid Butt\authorrefmark{1},
Giuseppe Caire\authorrefmark{2}, Ralf R.
M\"{u}ller\authorrefmark{1}}\\
\authorblockA{\authorrefmark{1}Institute of Electronics and Telecommunications,
Norwegian University of Science and Technology, Trondheim, Norway}
\authorblockA{\authorrefmark{2}Department of Electrical Engineering,
University of Southern California, Los Angeles, CA}
}

\maketitle

\begin{abstract}
We find the capacity region of linear finite-field deterministic networks
with many sources and one destination.
Nodes in the network are subject to interference and broadcast constraints,
specified by the linear finite-field deterministic model.
Each node can inject its own information as well as relay other
nodes'  information. We show that the capacity region coincides with the
cut-set region. Also, for a specific case of correlated sources we provide
necessary and sufficient conditions for the sources transmissibility.
Given the ``deterministic model'' approximation for the corresponding Gaussian
network model, our results may be relevant to wireless sensor networks where
the sensing nodes multiplex the relayed data from the other nodes with their own data,
and where the goal is to decode all data at a single ``collector'' node.
\end{abstract}

\section{Introduction} \label{intro}

Wireless sensor networks (WSN) consist of many
nodes with sensing, computation, and communication capabilities, sharing a common
wireless communication channel.
In a typical WSN configuration, a large number of nodes
measure possibly correlated data and transmit to a single collector node.
This network problem is referred to as the ``sensor reachback problem''
in \cite{servetto-barros}. In many applications, nodes are energy-limited
and the physical distance between each sensing node
and the common destination
makes the transmission difficult or (energy wise) expensive.
We investigate such a scenario where communicating nodes cooperate
with each other and act as relays in order to transport their own data along with the
data from the other sensing nodes.

Wireless channels differ from their wired line counterpart in two
fundamental aspects. On one hand, the wireless channel is a
broadcast (shared) medium and the signal from any transmitter is
received by potentially many receivers. This is called
\emph{broadcast} constraint.
On the other hand,  any receiver observes the superposition (linear combination)
of signals from possibly many transmitters. This is called \emph{interference}
constraint. The simultaneous presence of these two constraints makes
a general wireless network quite difficult to analyze. The multiuser
Gaussian channel that models a relay network, unfortunately, has so
far escaped a sharp general characterization, even in the simplest case of a
Gaussian relay network with a single source, single destination and
a single relay \cite{meulen}. The capacities of Gaussian relay channel and certain
discrete relay channels are evaluated in \cite{cover1} and a lower
bound to the capacity of general relay channel is presented. In
\cite{Gastpar}, capacity is determined for a Gaussian relay network
when the number of relays is asymptotically large.

In \cite{salman1}, a simpler deterministic channel is
proposed. While this channel model is significantly simpler to analyze, it is able to
capture the key aspects of broadcast and interference constraints.
For this model, referred to as the linear finite-field deterministic model,
\cite{salman1} determines the capacity for a general relay network with one source and one destination,
as well as the multicast capacity with one source, multiple destinations and common
information only. Our contribution in this paper builds heavily on the results and techniques
of \cite{salman1} and can be regarded almost as a trivial extension thereof.
Nevertheless, to the best of our knowledge and somehow surprisingly,
this simple extension has not been reported before.

We consider the ``sensor reachback problem'' \cite{servetto-barros}
for a linear finite-field deterministic network with arbitrary topology,
a single destination node and independent information at the source nodes.
We show that the capacity region for this network is given by the cut-set
bound and takes on a very simple and appealing closed-form expression.
Also, for a specific sources correlation model, we find necessary and sufficient conditions
for the sources transmissibility. This result reminds closely Theorem 1 of \cite{servetto-barros},
with the following main differences: on one hand, the result of \cite{servetto-barros} is more general since it applies to
general correlated discrete sources observed at the sensor nodes and general noisy channels.
On the other hand, our result applies to networks with broadcast and interference constraints while
the result of \cite{servetto-barros} requires ``orthogonal'' channels, i.e.,
with neither broadcast nor interference constraints.

We expect that the achievability technique for the Gaussian (noisy) relay network
proposed in \cite{salman1} can be generalized to the case of multiple independent sources
and a single destination as examined in this paper, so that a scheme that achieves a bounded and fixed gap to
the capacity region in the Gaussian case can be found.
Also, we believe that a fixed-gap rate-distortion achievable region can be
found using independent quantization and Slepian-Wolf binning for the
case of correlated Gaussian sources with mean-squared distortion and Gaussian noisy channels,
at least for some specific source correlation model (see \cite{Maddah-tse}), especially matched to
the discrete correlated source  model considered here.  This, however, seems to be a far more involved result
since event in the standard case of Gaussian/quadratic separated lossy encoding
(that corresponds to the case where the communication network reduces to a set of orthogonal links from the sensor nodes to
the destination),  a general fixed-gap characterization of the rate-distortion region is missing \cite{Maddah-tse}.

In this work we limit ourselves to the linear finite-field deterministic model
and we leave the fixed-gap achievability for the Gaussian case
to future work.

\section{Review of the deterministic linear finite-field model} \label{review}

In this section we briefly review the deterministic channel model proposed
in \cite{salman1} and used in this work. The received signal at each node is a deterministic function
of the transmitted signal. This model focuses on the signals interaction rather than on the channel noise.
In a Gaussian (real) network, a single link from node $i$ to node $j$ with SNR $\SNR_{i,j}$ has capacity
$C_{i,j} = \frac{1}{2} \log(1 + \SNR_{i,j}) \approx \log \sqrt{\SNR_{i,j}}$. Therefore, approximately,
$n_{i,j} =  \left \lceil \log \sqrt{\SNR_{i,j}} \right \rceil$
bits per channel use can be sent reliably. In \cite{salman1} (see also references therein),
the Gaussian channel is replaced by a finite-field deterministic model that reflects the
above behavior. Namely, the transmitted signal amplitude is represented through its
binary\footnote{The generalization to $p$-ary expansion is trivial. Here we focus on the binary expansion
as in \cite{salman1}.}
expansion $X = \sum_{\ell=1}^\infty B_\ell 2^{-\ell}$ where $B_\ell \in \FF_2$.
At the receiver, all the input bits such that
$\sqrt{\SNR_{i,j}} 2^{-\ell} > 1$ (i.e., received ``above the noise level'') are perfectly decoded, while all those such that
$\sqrt{\SNR_{i,j}} 2^{-\ell} \leq 1$ (i.e., received ``below the noise level'') are completely lost.
It follows that only the most significant bits (MSBs) can be reliably decoded, such that the capacity of the
deterministic channel is given exactly by $n_{i,j}$ and it is achieved by
letting $B_1, \ldots, B_{n_{i,j}}$ i.i.d. Bernoulli-$1/2$.

A linear finite-field deterministic relay network is defined as a directed acyclic
graph $\Gc = \{\Vc, \Ec\}$ such that the received signal at any node $j \in \Vc$ is given by
\begin{equation} \label{mac-channel}
\yv_j = \sum_{i \in \Vc : (i,j) \in \Ec} \Sm^{q - n_{i,j}} \xv_i
\end{equation}
where $\yv_j, \xv_i \in \FF_2^q$, sum and products are defined over the vector space $\FF_2^q$,
and where
\[ \Sm = \left [ \begin{array}{ccccc}
0 & 0  & 0 &\cdots &  0 \\
1 & 0  & 0 &\cdots &  0 \\
0 & 1  & 0 &\cdots &  0 \\
\vdots & \ddots & \ddots & \ddots & \vdots \\
0 & \cdots & 0 & 1 & 0 \end{array} \right ] \]
is a ``down-shift'' matrix. Notice that $n_{i,j} \leq q$ indicates the deterministic channel capacity
for the link $(i,j)$ as described before.
Without loss of generality, the integer $q$ can be set equal to the maximum of all $\{n_{i,j} : (i,j) \in \Ec\}$.
The broadcast constraint is captured by the fact that the input $\xv_i$ for each node $i$ is common to all
channels $(i,j) \in \Ec$.

In the case of single source (denoted by $s$) single destination (denoted by $d$), Theorem 4.3 of \cite{salman1} yields the capacity of
linear finite-field deterministic relay networks in the form
\begin{equation} \label{salman-capacity}
C = \min_{(\Sc, \Sc^c) \in \Lambda_d} \; {\rm rank} \left \{ \Gm_{\Sc,\Sc^c} \right \}
\end{equation}
where $\Lambda_d$ is the set of cuts $\Sc \subset \Vc$, $\Omega^c = \Vc - \Sc$ such that
$s \in \Sc$ and $d \in \Sc^c$, and where $\Gm_{\Sc,\Sc^c}$ is the transfer matrix for the cut
$(\Sc,\Sc^c)$, formally defined as follows. Let $\Nc(i)$ denote the set of nodes $j$ for which $(i,j) \in \Ec$
(this is the ``fan-out'' of node $i$) and  let $\Pc(j)$ denote the set of nodes $i$ for which $(i,j) \in \Ec$ (this is the ``fan-in'' of node $j$).
The transfer matrix $\Gm_{\Sc,\Sc^c}$ is defined as the matrix of the linear transformation
between the transmitted vectors (channel inputs) of nodes $\beta_{\rm in}(\Sc)$ and the received vectors (channel outputs) of nodes
$\beta_{\rm out}(\Sc)$, where the inner and outer boundaries $\beta_{\rm in}(\Sc)$ and $\beta_{\rm out}(\Sc)$ of
$\Sc$ are defined as \cite{kramer-note}:
\[ \beta_{\rm in}(\Sc) = \{ i \in \Sc : \Nc(i) \cap \Sc^c \neq \emptyset \} \]
and
\[ \beta_{\rm out}(\Sc) = \{ j \in \Sc^c : \Pc(j) \cap \Sc \neq \emptyset \} \]
In words: $\beta_{\rm in}(\Sc)$ is the set of nodes of $\Sc$ with a direct link to nodes in $\Sc^c$, and $\beta_{\rm out}(\Sc)$
is the set of nodes in $\Sc^c$ with a direct link from nodes in $\Sc$.

Going through the proof of Theorem 4.3 in \cite{salman1} we notice that the ``down-shift'' structure for the individual channels
is irrelevant. In fact, this structure is useful in making the connection between the linear finite-field model and the corresponding
Gaussian case. As a matter of fact, if the channel matrices $\Sm^{q - n_{i,j}}$ in the above model are replaced by general
matrices $\Sm_{i,j} \in \FF_2^{q \times q}$, the result (\ref{salman-capacity})
still holds.

\section{Main result} \label{main}

In a linear finite-field deterministic network defined as above,
let $\Vc = \{1,\ldots, N, d\}$, where node $d$ denotes the common destination and all other nodes
$\{1, \ldots, N\}$ have independent information to send to node $d$.
For any integer $T = 1, 2, \ldots$ we let $\Wc_i = \{1, \ldots, \lceil 2^{TR_i} \rceil \}$ denote
the message set of node $i = 1,\ldots, N$. A $(T, R_1, \ldots, R_N)$ code for the network is defined
by a sequence of {\em strictly causal} encoding functions
$f_i^{[t]} : \Wc_i \times \FF_2^{q(t - 1)} \rightarrow \FF_2^q$,
for $t = 1, \ldots, T$ and $i = 1, \ldots, N$,  such that the transmitted signal of node $i$ at (discrete) time $t$
is given by $\xv_i[t] = f_i^{[t]}(w_i, \yv_i[1], \ldots, \yv_i[t-1])$, and by a decoding function
$g : \FF_2^{Tq} \rightarrow \Wc_1 \times \cdots \Wc_N$, such that the set of decoded messages is given by
$(\widehat{w}_1, \ldots, \widehat{w}_N) = g(\yv_d[1], \ldots, \yv_d[T])$.

The average probability of error for such code is defined as
$P_n(e) = \PP((W_1, \ldots, W_N) \neq (\widehat{W}_1, \ldots, \widehat{W}_N)$,
where the random variables $W_i$ are independent and uniformly distributed on the corresponding message
sets $\Wc_i$. The rate $N$-tuple $(R_1, \ldots, R_N)$ is {\em achievable}
if there exists a sequence of $(T, R_1, \ldots, R_N)$-codes with $P_n(e) \rightarrow 0$ as
$T \rightarrow \infty$. The capacity region $\Cc$ of the network is the closure of the set of all
achievable rates. With these definitions, we have:

\begin{thm} \label{thm1}
The capacity region $\Cc$ of a linear finite-field deterministic network $(\Vc, \Ec)$ with
independent information at the nodes $\{1,\ldots, N\}$ and a single destination $d$ is given by
\begin{equation} \label{cut-set-general}
\sum_{i \in \Sc} R_{i} \leq {\rm rank}\left \{ \Gm_{\Sc, \Sc^c} \right  \}, \;\;\; \forall \; \Sc \subseteq \{1,\ldots, N\}.
\end{equation}
\end{thm}

\begin{proof}
The converse of (\ref{cut-set-general}) follows directly from the general cut-set bound
and by the fact that, for the linear deterministic network model, uniform i.i.d. inputs maximize
all cut-set values at once \cite{cover_book,salman1,kramer-note}.

For the direct part, we build an augmented network by introducing a virtual source node 0 and by expanding
the channel output alphabet of each node $i  = \{1, \ldots, N\}$. Let $\{n_{0,i} :  i = 1,\ldots, N\}$ be arbitrary
non-negative integers. The channel output alphabet of node $i$ in the augmented network is given by
$\FF_2^{q + n_{0,i}}$. The virtual source node 0 has $n_0 = \sum_{i=1}^N n_{0,i}$ input bits,
partitioned into $N$ disjoint sets $\Uc_i$ of cardinality $n_{0,i}$ for $i = 1,\ldots, N$,
respectively, such that the bits of subset $\Uc_i$ are sent directly to node $i$ and are
received at the top $n_{0,i}$ MSB positions of the expanded channel output alphabet.
Fig. \ref{fig:diamond_aug} shows an example of such network augmentation for a ``diamond'' network \cite{salman1}.

\begin{figure}[!htb]
 \centering
 \includegraphics[width=3.0 in]{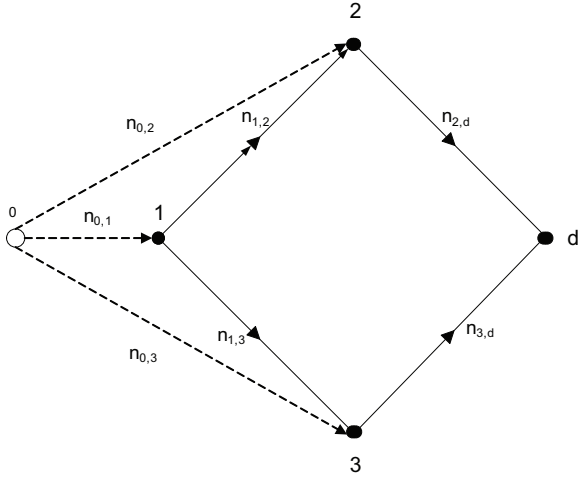}
  \caption{A diamond network with a source node $1$, two relay nodes $2$ and $3$ and a common destination $d$ is augmented
by adding node $0$ and virtual links to nodes $1$, $2$ and $3$.}
  \label{fig:diamond_aug}
\end{figure}

After introducing the virtual source node, the augmented linear finite-field deterministic network belongs to the class
studied in \cite{salman1} with the minor difference that the channel linear transformations are not necessarily
limited to ``down-shifts''. Nevertheless, as we observed before,  Theorem 4.3 of  \cite{salman1} still applies. Letting
$R_0$ denote the rate from the virtual source node 0 to the destination node $d$, we have that all rates $R_0$ satisfying
\begin{equation} \label{suca}
R_0 \leq \min_{(\Omega_0 , \Omega_0^c) \in \Lambda_d} \;\; {\rm rank} \left \{ \Gm_{\Omega_0,\Omega_0^c} \right \}
\end{equation}
are achievable,  where $\Lambda_d$ is the set of all cuts $(\Omega_0, \Omega_0^c)$ of the augmented network such that
$0 \in \Omega_0$ and $d \in \Omega_0^c$.

For any such set $\Omega_0$ we have that $\Omega_0 = \Sc \cup \{0\}$, for some
$\Sc \subseteq \{1, \ldots, N\}$.  Consequently, we have that $\Omega_0^c = \Sc^c$, where $\Sc, \Sc^c$
are subsets as defined in the statement of Theorem \ref{thm1}.
Since the links from 0 to any nodes $i \in \{1,\ldots, N\}$ are orthogonal by construction
(not subject to any broadcast or interference constraint), we have  that $\Gm_{\Omega_0,\Omega_0^c}$
has a block-diagonal form where a block is given by  $\Gm_{\Sc, \Sc^c}$ (the links of the original network, corresponding to the cut
$(\Omega_0,\Omega_0^c)$ via the correspondence  $\Omega_0 \leftrightarrow \Sc$ defined above)
and other blocks, denoted by $\Gm_{0,j}$ for all $j \in \Sc^c$, have rank $n_{0,j}$, respectively.
By construction, there is no direct link between $0$ and $d$ so, without loss of generality, we can assume
$n_{0,d} = 0$.  The general form for $ \Gm_{\Omega_0,\Omega_0^c}$ is
\[  \Gm_{\Omega_0,\Omega_0^c} = \left [ \begin{array}{cccc}
 \Gm_{\Sc,\Sc^c} & 0 & \cdots  & 0 \\
 0 & \Gm_{0,i_1}  &                  & \vdots \\
 \vdots &               & \ddots &   0 \\
 0 &   \cdots   &    0 & \Gm_{0,i_|\Sc^c|} \end{array} \right ] \]
where we have indicated $\Sc^c = \{i_1, \ldots, i_{|\Sc^c|}\}$.  Therefore, we have
\begin{equation} \label{suca1}
{\rm rank} \left \{ \Gm_{\Omega_0,\Omega_0^c} \right \} = {\rm rank} \left \{ \Gm_{\Sc,\Sc^c} \right \} + \sum_{j \in \Sc^c} n_{0,j}
\end{equation}
In particular, the cut $\Omega_0 = \{0\}$ yields
\begin{equation} \label{suca2}
R_0 \leq \sum_{j=1}^N n_{0,j}
\end{equation}
By letting this inequality hold with equality, and by replacing this into all other inequalities,
we obtain the set of inequalities
\begin{equation} \label{suca3}
\sum_{i \in \Sc} n_{0,i} \leq  {\rm rank} \left \{ \Gm_{\Sc,\Sc^c} \right \}, \;\;\;\; \forall \;\; \Sc \subseteq \{1, \ldots, N\}
\end{equation}
where we used the fact that $\sum_{j=1}^N n_{0,j} - \sum_{j \in \Sc^c} n_{0,j} = \sum_{i\in \Sc} n_{0,i}$.

Consider now the ensemble of augmented networks for which there exist integers $\{n_{0,i} : i = 1, \ldots,N \}$
that satisfy (\ref{suca3}). For such networks, the rate $R_0 = \sum_{j=1}^N n_{0,j}$ is
achievable (by \cite{salman1}) and therefore the individual rates $R_i = n_{0,i}$ are achievable
by the argument above. Finally, the closure of the convex hull of all individual rate vectors $\Rm = (n_{0,1}, \ldots, n_{0,N})$ of such networks is achievable by time-sharing. It is immediate to see that this convex hull is provided by  the inequalities (\ref{cut-set-general}).\footnote{Indeed, the inequalities (\ref{cut-set-general}) represent the convex relaxation of the integer constraints (\ref{suca2}).}
\end{proof}

\section{A specific example: diamond network} \label{diamond}

In this section we work out a simple example and provide an explicit
achievability strategy.  Consider the ``diamond'' network shown in Fig.
\ref{fig:diamond_aug}, with nodes $\{1,2,3,d\}$ and links of
capacity $n_{1,2}, n_{1,3}, n_{2,d}$ and $n_{3,d}$.
In this case, Theorem \ref{thm1} yields the capacity region $\Cc$ given by
\begin{eqnarray} \label{eqn:diamond-cap}
R_1 + R_2 + R_3 & \leq & \max\{ n_{2,d}, n_{3,d}\}  \\
R_1 + R_2 & \leq & n_{2,d} + n_{1,3}  \\
R_1 + R_3 & \leq & n_{3,d} + n_{1,2}  \\
R_1 & \leq & \max\{ n_{1,2}, n_{1,3}\}  \\
R_2 & \leq & n_{2,d} \\
R_3 & \leq & n_{3,d}
\end{eqnarray}

Next, we provide simple coding strategies that achieve all relevant
vertices of $\Cc$. Any point $\Rm \in \Cc$ can be obtained by
suitable time-sharing of the vertices-achieving strategies. There
are 24 possible orderings of the individual link capacities
$n_{1,2}, n_{1,3}, n_{2,d}$ and $n_{3,d}$. Due to symmetry, the
regions for the case $n_{3,d} > n_{2,d}$ will be the mirror image of
the regions for the case $n_{2,d} > n_{3,d}$. Therefore, we shall
consider only the cases where $n_{2,d} \geq n_{3,d}$.

\begin{figure}[!htb]
 \centering
 \includegraphics[width=8cm,height=6cm]{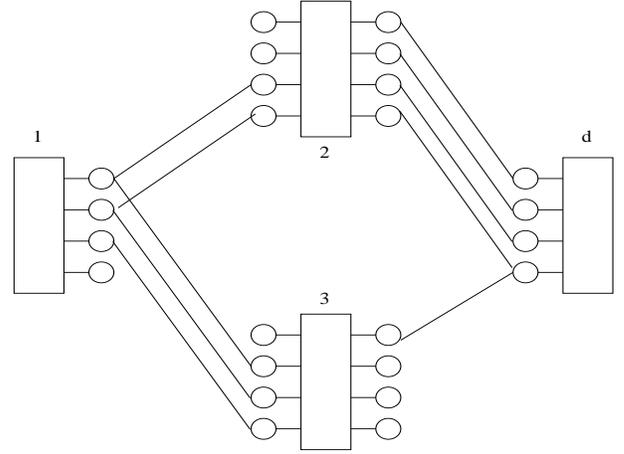}
  \caption{The configuration of the diamond network in the example (Case (1) in Fig.\ref{region1})}.
  \label{diamond-1234}
\end{figure}

\begin{figure}[!htb]
 \centering
 \includegraphics[width=8cm]{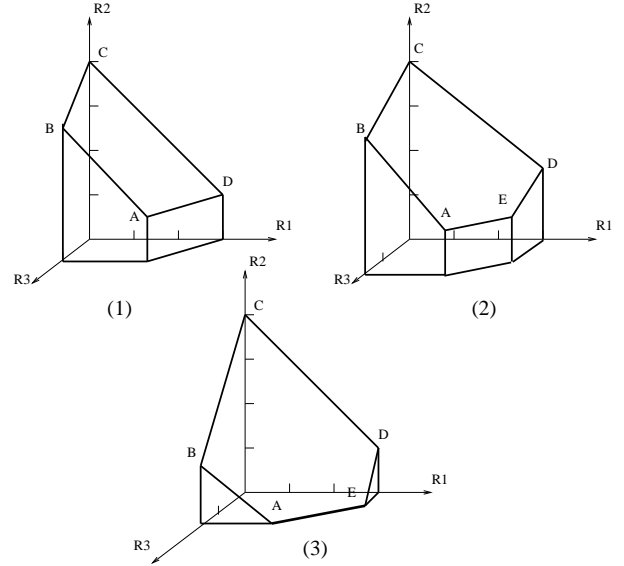}
  \caption{The capacity region of the diamond network in the example.}
  \label{region1}
\end{figure}

The remaining 12 cases have to be discussed individually.  For
example, let's focus on the case $n_{3,d} \leq n_{1,2} \leq n_{1,3}
\leq n_{2,d}$. An example of the network for the choice of the link capacities
$n_{3,d}=1, n_{1,2}=2, n_{1,3}=3, n_{2,d}=4$ is given in Fig.
\ref{diamond-1234}.
Fig. \ref{region1} shows qualitatively the shape of the capacity region in
the three possible sub-cases of the link-capacity
ordering $n_{3,d} \leq n_{1,2} \leq n_{1,3} \leq n_{2,d}$:
case 1) for $n_{1,2} + n_{3,d} < n_{1,3}$; case 2) for $n_{1,2} + n_{3,d} \geq
n_{1,3}$, and case 3) for $n_{1,2} + n_{3,d} \geq n_{2,d}$.
In all cases, the achievability of the vertices B and C of the region
of Fig. \ref{region1} is trivial, since these correspond to vertices
of the multi-access channel with node 2 and 3 as transmitters and
node $d$ as receiver.

{\bf Case 1).} Vertex A has coordinates  $(R_1=n_{1,2},
R_2=n_{2,d} - n_{1,2} - n_{3,d} , R_3= n_{3,d})$ and can be achieved
by letting node 1 send $n_{1,2}$ to node 2. Node 2 decodes and
forwards these bits after multiplexing its own $n_{2,d} - n_{1,2} -
n_{3,d} > 0$ bits in the MSB positions, such that node 3 can send
$n_{3,d}$ bits without interference from node 2.
Vertex D has coordinates  $(R_1=n_{1,2} + n_{3,d}, R_2=n_{2,d} - n_{1,2} -
n_{3,d} , R_3= 0)$ and can be achieved by letting node 1 send
$n_{1,2} + n_{3,d}$ bits. These can be all
decoded by node 3, then node 3 can forward the bottom
(least-significant) $n_{3,d}$ bits of node 1 to node $d$.
Node 2 decodes the top (most-significant) $n_{1,2}$
bits from node 1, and forwards them
after multiplexing its own bits.

{\bf Case 2).} Vertices A, D and E have coordinates $(R_1=n_{1,2},
R_2=n_{2,d} - n_{1,2}-n_{3,d}, R_3= n_{3,d})$, $(R_1=n_{1,3},
R_2=n_{2,d} - n_{1,3}, R_3=0)$ and $(R_1=n_{1,3}, R_2=n_{2,d} -
n_{1,2}-n_{3,d}, R_3= n_{1,2}+n_{3,d}-n_{1,3})$, respectively.
Vertex A can be achieved in the same way as in Case 1).
Vertex D can be achieved by letting node 1 send $n_{1,3}$ bits to node 3.
Node 3 decodes and forwards the bottom $n_{3,d}$. Since in this case
$n_{1,2} \geq n_{1,3} - n_{3,d}$, node 2 can decode the top $n_{1,3} - n_{3,d}$ bits of node 1,
and forwards them to node $d$ after multiplexing its own $n_{2,d} - n_{1,3}$ bits, using its $n_{2,d} - n_{3,d}$ MSBs.
Vertex E can be achieved by letting node 1 transmit $n_{1,3}$ bits, where the top $n_{1,2}$ of which are received by node 2.
Node 3 forwards the bottom $n_{1,3} - n_{1,2}$ bits of node 1, and multiplex its own $n_{3,d} + n_{1,2} - n_{1,3}$ bits.
Node 2 forwards the top $n_{1,2}$ bits from node 1, by multiplexing its own $n_{2,d} - n_{1,2} - n_{3,d}$ bits, transmitting
over its $n_{2,d} - n_{3,d}$ MSBs.

{\bf Case 3).} Vertices A, D and E have coordinates
$(R_1=n_{2,d}-n_{3,d}, R_2=0, R_3= n_{3,d})$,
$(R_1=n_{1,3},R_2=n_{2,d} - n_{1,3}, R_3=0)$ and $(R_1=n_{1,3},
R_2=0, R_3=n_{2,d}-n_{1,3})$, respectively. Vertex A can be achieved
by letting node 1 send $n_{2,d} - n_{3,d}$ bits to node 2. Since
$n_{2,d} - n_{3,d} \leq n_{1,2}$ these can be decoded and forwarded
to node $d$ in the MSB positions. Node 3 simply sends $n_{3,d}$ bits
to node $d$ without interfering with node 2. Vertex D is achieved by
letting node 1 send $n_{1,3}$ bits. The top $n_{1,3} - n_{3,d}$ of
these are decoded by node 2 and forwarded together with $n_{2,d} -
n_{1,3}$ own bits. The bottom $n_{3,d}$ bits of node 1 are decoded
and forwarded by node 3. Finally, vertex E is achieved by letting
node 1 send $n_{1,3}$ bits. The bottom $n_{3,d} - n_{2,d} + n_{1,3}$
of these are forwarded by node 3, after multiplexing its own
$n_{2,d} - n_{1,3}$ bits. Since $n_{2,d} - n_{3,d} \leq n_{1,2}$,
node 2 can decode the top $n_{2,d} - n_{3,d}$ bits from node 1 and
forward them to node $d$ using its MSB positions. Other cases follow
similarly and the whole capacity region is achieved by decode and
forward.

\section{Transmissibility for correlated sources} \label{correlated}

Consider the case of a sensor network where the nodes $\{1, \ldots, N\}$ observe samples from a spatially-correlated,
i.i.d. in time, discrete vector source $\Um = (U_1, \ldots, U_N)$ (see the source model in \cite{servetto-barros}).
The goal is to reproduce the source blocks $\uv[1], \ldots, \uv[T]$ at the common destination node $d$.
If the source blocks can be recovered at the destination with vanishing probability of error
as $T \rightarrow \infty$, the vector source is said to be {\em transmissible}.
In the case of a network of orthogonal links with capacities $C_{i,j}$,
this problem was solved in \cite{servetto-barros} and yields the necessary and sufficient
transmissibility condition\footnote{The notation $U_\Sc = \{U_i : i \in \Sc\}$ is standard.}
\begin{equation} \label{servetto-result}
H(U_\Sc|U_{\Sc^c}) \leq \sum_{(i,j) \in \Sc\times \Sc^c} C_{i,j}, \;\;\; \forall \; \Sc \subseteq \{1, \ldots, N\}.
\end{equation}
From the system design viewpoint, the above result yields the optimality of the
``separation'' approach consisting of the concatenation of Slepian-Wolf coding for the source with
routing and single-user channel coding for the network \cite{servetto-barros}.

With the same assumptions and linear finite-field deterministic network defined before,
we consider a specific model for the vector source as defined in \cite{Maddah-tse}.
Let $n_0$ be a non-negative integer, and let $\Vm \in \FF_2^{n_0}$ be a random vector of
uniform i.i.d. bits. For all $i = 1, \ldots, N$, let $\Uc_i \subseteq \{1, \ldots, n_0\}$ and
define $U_i \in \FF_2^{|\Uc_i|}$ as the restriction of $\Vm$
to the components $\{V_\ell : \ell \in \Uc_i\}$ of $\Vm$. Then, the correlation model for the source $(U_1,\ldots, U_N)$ is reduced to
the following ``common bits'' case:
sources $U_i$ and $U_j$ have common part $\{V_\ell : \ell \in \Uc_i \cap \Uc_j\}$
while the bits $V_\ell$ in $\Uc_i - \Uc_j$ and in $\Uc_j - \Uc_i$  are mutually independent.
It follows that $H(U_i|U_j) = |\Uc_i| - |\Uc_i \cap \Uc_j|$.

This source model is somehow ``matched'' to a correlated source defined over the reals in the following intuitive sense.
Consider $N = 2$ and let $U_1$ and $U_2$ denote the binary quantization indices resulting from quantizing two correlated random variables
$A_1 \in \RR$ and $A_2 \in \RR$ using ``embedded'' scalar uniform quantizers
with $n$ bits, such that their first $m$ MSBs are identical and their last $n-m$ least significant bits (LSBs) are mutually
independent. If $A_1, A_2$ are marginally uniform and symmetric, $U_1$ and $U_2$ are {\em exactly}
obtained by defining $\Vm$ as above, with $n_0 = 2n - m$ independent bits, and letting $U_1$  include the $m$ MSBs and the
first set of $n - m$ LBSs of $\Vm$, and $U_2$ include the same $m$ MSBs and the
second set of $n - m$ LBSs of $\Vm$. This model trivially generalizes to the case of $N$ correlated sources and is related to
the Gaussian sources with ``tree'' dependency considered in \cite{Maddah-tse}.
For the source model defined above we have the following simple result:

\begin{thm} \label{thm2}
The vector source $\Um = (U_1,\ldots, U_N)$ is transmissible over the linear finite-field deterministic
network $(\Vc, \Ec)$  if and only if
\begin{equation} \label{servetto-linear-ff}
H(U_\Sc|U_{\Sc^c}) \leq {\rm rank}\left \{ \Gm_{\Sc, \Sc^c} \right  \}, \;\;\; \forall \; \Sc \subseteq \{1,\ldots, N\}.
\end{equation}
\end{thm}

\begin{proof}
Again, we consider an augmented network with a single source node denoted by $0$,
with $n_0$ output bits that we denote by $\Vm$. As before, subsets $\Uc_i$ of
cardinalities $n_{0,i}$ of these bits are sent to nodes $i$, respectively. However, differently from before
we choose the subsets $\Uc_i$ to overlap in accordance with the vector source model.
For the augmented network, the rate $R_0$ from the virtual source to the destination $d$
must satisfy (\ref{suca}). In particular, choosing $\Omega_0 = \{0\}$ we get $R_0 \leq n_0$.
Generalizing the proof of Theorem \ref{thm1} to the case of overlapping sets $\{\Uc_i\}$,
we find that for any cut $(\Omega_0, \Omega_0^c)$ of the augmented network such that
$\Omega_0 = \Sc \cup \{0\}$ and $\Omega_0^c = \Sc^c$, with $\Sc \subseteq \{1, \ldots, N\}$ we have
\[ {\rm rank} \left \{ \Gm_{\Omega_0,\Omega_0^c} \right \} =
{\rm rank} \left \{ \Gm_{\Sc,\Sc^c} \right \} + {\rm rank} \left \{ \Gm_{0,\Sc^c} \right \} \]
where $\Gm_{0,\Sc^c}$ is the linear transformation between the inputs $\Vm$ and the (augmented) channel outputs
of nodes $j \in \Sc^c$. By construction, the matrix $\Gm_{0,\Sc^c}$ is formed by linear independent columns
for all bits $V_\ell$ with $\ell \in \bigcup_{j \in \Sc^c} \Uc_j$. Therefore,
\[ {\rm rank} \left \{ \Gm_{0,\Sc^c} \right \} = \left |\bigcup_{j \in \Sc^c} \Uc_j \right | = H(U_{\Sc^c}) \]
Since $\Vm$ is uniform i.i.d., we have $R_0 = n_0 = H(\Vm) = H(\Um)$.
Replacing these equalities into the set of inequalities (\ref{suca}) and using the chain rule of entropy
$H(\Um) = H(U_\Sc|U_{\Sc^c}) + H(U_{\Sc^c})$ we obtain
that the conditions (\ref{servetto-linear-ff}) are sufficient for transmissibility.
On the other hand, if a source as defined in our model was transmissible, then the set of conditions (\ref{servetto-linear-ff}) must
hold, otherwise the rate $R_0$ of the corresponding single-source single destination augmented network would violate
(\ref{suca}). Hence, necessity also holds.
\end{proof}

\section{Conclusions \label{conclusions}}

In this work we have characterized the capacity region for a linear finite-field deterministic network
with independent information at all nodes and a single destination node.
In our setup, all nodes may relay information from other nodes as well as inject their own
information into the network.
This may serve as a simplified model for a large WSN where sensing nodes cooperate with each
other to send the collective data towards a single collector node.
For a specific model of discrete binary source correlation at the nodes, we have also found necessary and sufficient
conditions for the source transmissibility.
Albeit restrictive, this correlation model may be useful (e.g., see \cite{Maddah-tse}) as a simple discrete ``equivalent''
(up to some bounded mean-square distortion penalty) for a spatially-correlated real sources whose components are observed
and encoded separately at the network nodes.

Motivated by these results, it is natural to investigate the performance of
achievability schemes based on the techniques as in \cite{salman1} (for independent information)
and separated quantization and Slepian-Wolf binning (for lossy transmission of correlated sources) in order to achieve the capacity region or the distortion region of actual WSN, within a bounded performance gap.

\bibliographystyle{IEEEtran}
\bibliography{Literatur}

\end{document}

%% file: macros.tex
\long\def\comment#1{}

\newfont{\bbb}{msbm10 scaled 700}

\newfont{\bb}{msbm10 scaled 1100}

\newcommand{\PP}{\mbox{\bb P}}
\newcommand{\RR}{\mbox{\bb R}}

\newcommand{\FF}{\mbox{\bb F}}

\newcommand{\uv}{{\bf u}}

\newcommand{\xv}{{\bf x}}
\newcommand{\yv}{{\bf y}}

\newcommand{\Gm}{{\bf G}}

\newcommand{\Rm}{{\bf R}}
\newcommand{\Sm}{{\bf S}}

\newcommand{\Um}{{\bf U}}

\newcommand{\Vm}{{\bf V}}

\newcommand{\Cc}{{\cal C}}

\newcommand{\Ec}{{\cal E}}

\newcommand{\Gc}{{\cal G}}

\newcommand{\Nc}{{\cal N}}

\newcommand{\Pc}{{\cal P}}

\newcommand{\Sc}{{\cal S}}

\newcommand{\Uc}{{\cal U}}
\newcommand{\Wc}{{\cal W}}
\newcommand{\Vc}{{\cal V}}

\newcommand{\SNR}{{\sf snr}}